\documentclass[11pt, conference]{IEEEtran}
\usepackage{amsfonts,amsmath,amssymb}

\newtheorem{theorem}{Theorem}[section]

\begin{document}

\title{A Framework for Partial Secrecy}

\author{
\authorblockN{Paul Cuff}
\authorblockA{Department of Electrical Engineering \\
Princeton University\\
E-mail: cuff@princeton.edu }
}

\maketitle

\begin{abstract}
We consider theoretical limits of partial secrecy in a setting where an eavesdropper attempts to causally reconstruct an information sequence with low distortion based on an intercepted transmission and the past of the sequence.  The transmitter and receiver have limited secret key at their disposal but not enough to establish perfect secrecy with a one-time pad.  From another viewpoint, the eavesdropper is acting as an adversary, competing in a zero-sum repeated game against the sender and receiver of the secrecy system.  In this case, the information sequence represents a sequence of actions, and the distortion function captures the payoff of the game.

We give an information theoretic region expressing the tradeoff between secret key rate and max-min distortion for the eavesdropper.  We also simplify this characterization to a linear program.  As an example, we discuss how to optimally use secret key to hide Bernoulli-$p$ bits from an eavesdropper so that they incur maximal Hamming distortion.
\end{abstract}

\section{Introduction}
\label{section introduction}

Secret communication by means of basic ciphers and steganography has been used since antiquity.  By the time of World War II, mechanical cipher machines such as Germany's Enigma brought cryptography to an advanced level.  Cipher systems were designed with secret and complex construction and often made use of secret keys.  As new systems were being developed, cryptanalysts employed mathematics and espionage to break the secrecy of such systems, with substantial success.

The theoretical investigation of secret communication was formalized by Claude Shannon as a fundamental component of his broad theory of communication.  In his 1949 paper, ``Communication Theory of Secrecy Systems'' \cite{ShannonSecrecy49}, Shannon showed among other things that truly secret communication over a public channel cannot be achieved with less secret key than is needed for a Vernam cipher using a one-time pad.  The Vernam cipher utilizes shared secret key to hide the message via a bit-wise exclusive-or operation.  This produces a transmitted signal that is independent of the message if the key is an independent sequence of random bits, as with the one-time pad.  To formalize the study of secret communication, Shannon assumed that an eavesdropper knows everything about the encryption system except the secret key.  Furthermore, he allowed the eavesdropper to use unlimited computational resources to break the encryption.  These assumptions lay foundation for strong statements about secrecy.  However, the most famous result from this work is that perfect secrecy, where the ciphertext (transmission) reveals nothing about the plaintext (message), requires a secret key with as much entropy as the message itself.

Exchanging such large secret keys for perfect secrecy can be unrealistic.  Thus, the field has advanced through a number of different compromises:  computational limits on the eavesdropper; noisy public channels for communication; metrics for partial secrecy; etc.

Practical modern applications of cryptography have been mostly based on secrecy that is computationally difficult to discover.  Diffie and Hellman introduced in \cite{Diffie76newdirections} the use of asymmetric keys for cipher systems and trapdoor functions, which are easy to compute but hard to invert, to keep the keys secret.  Systems of this type maintain secrecy as long as the trapdoor function is not inverted for the particular key used.  To date, complexity theory has yet to provide certificates of proof that commonly used encryption systems are computationally difficult to break.  Nevertheless, in practice such systems have proven extremely useful.

Mathematical proofs of secrecy along the lines of Shannon have remained an active area of research.  Information theoretic secrecy may someday be a cost effective and desirable framework for encryption if an inexpensive method for secret key exchange is developed, which may come about through the use of quantum channels as pioneered by Bennett and Brassard \cite{BennettBrassard84}.  Additionally, quantum computing may bring security risks to the state-of-the-art cryptographic systems that don't provide information theoretic secrecy.  However, perhaps more importantly, the study of information theoretic secrecy builds a foundation of precise problems, with answers that provide important insights into the nature of secret communication.  These insights may be used to shape the development of practical systems for encryption in complex settings.

Since Shannon, various settings have been explored.  The wire-tap channel is a setting introduced by Wyner \cite{WynerWiretap75}, where the intended receiver and an eavesdropper each get different noisy versions of the transmitted signal.  If the eavesdropper has a noisier view of the signal than the intended receiver, then perfect secrecy can be achieved without the use of a secret key, although a low rate of communication might be required.  This has inspired related work on physical layer security.  Some settings allow for correlated (but not identically equal) measurements to be made by the various participants in consideration, from which they can derive a secret key or make inferences about the information source.  See, for example, \cite{4594958}.

Even in the simplest of settings, a general theory of secrecy must address partial secrecy.  The following section discusses this topic, which is the focus of this work.

\section{Partial Secrecy}

Most work on information theoretic security, including Shannon's \cite{ShannonSecrecy49} and Wyner's \cite{WynerWiretap75}, address the concept of imperfect or partial secrecy.  A widely used metric of partial secrecy is {\em equivocation}.  The term ``equivocation'' was used in an information theory context as early as Shannon's 1948 paper \cite{Shannon48}.  In the process of communicating through a noisy channel, the receiver of the communication may be confused (due to the noise) as to which signal was actually transmitted.  The goal of a reliable communication system is to remove the confusion.  For the purpose of providing an intuitive illustration, Shannon considered the analog transmission of a random information signal over a noisy channel.  The result is equivocation, in the English language sense of the word.  Because of the noise, the receiver cannot uniquely identify the meaning of the information, which is the source sequence that was transmitted.  Shannon quantified equivocation as the conditional entropy of the source sequence $X^n$ given the output of the channel $Y^n$.

In a secrecy system, equivocation of the information signal with respect to the eavesdropper's knowledge is a desirable outcome.  Let ${\cal E}$ represent everything that the eavesdropper knows.  If perfect secrecy is achieved, then the eavesdropper's knowledge, ${\cal E}$, is independent of the information, $X^n$.  In this case, the equivocation $H(X^n|{\cal E})$ is equal to its upper bound of $H(X^n)$.  On the other hand, if the equivocation is zero, then the eavesdropper knows the information perfectly.  Consequently, equivocation has been adopted repeatedly as a measurement of partial secrecy.  It is also a nice quantity to deal with for information theorists.  In a variety of work, including for example \cite{4594958}, the greatest possible equivocation is sought under the communication and secret key constraints.

Shannon referred extensively to equivocation in his landmark secrecy paper \cite{ShannonSecrecy49}.  He outlined the properties of equivocation that make it an intuitive and useful quantity to deal with.  However, the mathematical use of equivocation in his work served as an intermediate mechanism for determining how long a cipher can be used before it can be cracked with reasonable certainty.  In his setting, the roughly linear relationship between the equivocation of the secret key and the length of the ciphertext made the analysis of equivocation a useful avenue to the results.  The paper does not use equivocation as a metric of partial secrecy.

Equivocation is not a thoroughly motivated measurement for partial secrecy, but it does have some interesting consequences and properties.  For example, equivocation of a message implies a lower bound on the size of a list that an eavesdropper can reliably limit the message to.  If the eavesdropper has some other description of the information to supplement the intercepted transmission, he might be able to deduce the information in spite of the encryption, but the amount of additional information must be at least as great as the equivocation created by the encryption system.  In the setting of \cite{ShannonSecrecy49}, the extra side information comes implicitly from the redundancy of the information source, and hence the ability to compress.  For instance, if an eavesdropper is aware of a compression algorithm that is more efficient than the compression used by the encoder in an encryption system, then he can use this ability to rule out unlikely sequences from the information source.  This would be like an eavesdropper intercepting a partially encrypted video sequence and decoding a (exponentially large) list of candidate videos.  If the eavesdropper then painstakingly watches each video in the list, most will be ruled out as garbage.  The redundancy of the source effectively provides side information to the eavesdropper.

Other approaches for measuring and maximizing partial secrecy have been explored.  In the work of Merhav \cite{1176626}, secrecy is measured by the probability that an eavesdropper can guess the message correctly.  This uses large deviations analysis to evaluate and minimize that probability.  Under perfect secrecy, the probability that an eavesdropper guesses an i.i.d. information sequence correctly goes exponentially to zero.  In a different approach, Maurer \cite{Maurer93} suggested a combination of computational and information theoretic secrecy.  He proposed an encryption scheme that uses very little secret key and a very long public random sequence.  The secret key selects parts of the random sequence to combine and use as a one-time pad.  Assuming that the eavesdropper cannot reasonably explore all parts of the random sequence, a high probability event yields the message and the transmission conditionally independent.  Thus, with an assumed computational limit, the system yields {\em conditionally-perfect secrecy} with high probability.

We take a different approach to partial secrecy.  Just as rate-distortion theory generalizes lossless compression, we can generalize lossless encryption by assigning a distortion function to the eavesdropper.  Suppose the eavesdropper attempts to reconstruct the data.  We don't care whether the eavesdropper reconstructed the exact sequence, but instead we care about the average distortion that the eavesdropper incurred.  In the simplest case, which we confine ourselves to in this work, we require the intended receiver to decode the information losslessly (with high probability) and attempt to force the eavesdropper to have a poor reconstruction of the data.

Our work is similar in ambition to Yamamoto's approach in \cite{YamamotoCipherDistortion97}.  He also assigned a distortion function to the eavesdropper.  However, the related bounds presented in that paper miss a crucial point.  It turns out that it is too easy to force the eavesdropper to have a poor reconstruction.  With only a trickle of secret key at any positive rate, the eavesdropper will reconstruct as poorly as if there was perfect secrecy.  This result will not be proven or thoroughly discussed in this paper.  But needless to say, a more interesting notion of partial secrecy is desired for the sake of a general theory.

In contrast to Yamamoto's setting, in our work the eavesdropper is allowed to reconstruct the source causally, based on the intercepted transmission and the past information in the sequence, represented mathematically as
\begin{eqnarray*}
\mbox{Eavesdropper's estimate of } X_i & = & Z_i(I,X^{i-1}),
\end{eqnarray*}
where $I$ is the transmission.  In this case, an interesting notion of partial secrecy emerges.  We establish a simple tradeoff between the rate of secret key and the max-min distortion that the eavesdropper will incur.

Secrecy is naturally motivated by competitive or adversarial settings.  Here also we find ourselves in a competitive setting, where an eavesdropper tries to reconstruct information, and the transmitter does all within his power to make this difficult.  In fact, since the eavesdropper can see the past information symbols, and appropriate situation to apply this result is in a repeated game.  The eavesdropper sees the past actions in the game and dynamically responds with the next action $Z_i(I,X^{i-1})$.  The distortion function then becomes the payoff of the zero-sum game.

\section{Problem Statement}

Let the source $\{X_i\}_{i=1}^{\infty}$ be an i.i.d. random process where the distribution of each element of the sequence $X_i$ is given by the probability mass function $p_0(x)$.  Let ${\cal X}$ be the support of $X_i$.

An $(R_0, R, n)$ code consists of an encoding function and a decoding function, utilizing a secret key rate of $R_0$ bits per symbol and a description rate of $R$ bits per symbol.  Specifically,
\begin{eqnarray*}
f & : & {\cal X}^n \times [2^{n R_0}] \to [2^{n R}], \\
g & : & [2^{n R}] \times [2^{n R_0}] \to {\cal X}^n.
\end{eqnarray*}

The secret key is represented by $\omega \sim Unif[2^{n R_0}]$.  The message $I = f(X^n,\omega)$, and the reconstruction $\hat{X}^n = g(I,\omega)$.  The encoder and decoder need not be deterministic; however, randomized encoding and decoding is not needed in this setting.  The converse part of the proof in Section \ref{section converse} holds even for randomized encoders and decoders.

A distortion function $d(x,z)$ takes on finite values,
\begin{eqnarray*}
d & : & {\cal X} \times {\cal Z} \to \Re.
\end{eqnarray*}

We say the rate pair $(R_0,R)$ can force distortion $D$ if there exists a sequence of $(R_0,R,n)$ codes for $n=1,2,...$ such that
\begin{eqnarray*}
\mathbf{P} (\hat{X}^n = X^n) & \to & 1, \\
\frac{1}{n} \sum_{q=1}^n \min_{z^{(q)} (i,x^{q-1})} \mathbf{E} \; d(X_q,z^{(q)}(I,X^{q-1})) & \geq & D.
\end{eqnarray*}
The rate-distortion triple $(R_0,R,D)$ is {\em achievable} for secrecy with lossless transmission if the rate pair $(R_0,R)$ can force distortion $D$.

Let the secrecy rate-distortion region ${\cal S}$ be the closure of achievable $(R_0,R,D)$ triples.

\section{Main Result}

The secrecy rate-distortion region ${\cal S}$ is equal to ${\cal S}_0$ stated below, as specified in the Theorem \ref{theorem main result}.

Define
\begin{eqnarray*}
{\cal S}_0 & \triangleq & \left\{
\begin{array}{rcl}
(R_0,R,D) & : & \\
& \exists & p(x,u) = p_0(x)p(u|x) s.t. \\
R_0 & \geq & H(X|U), \\
R & \geq & H(X), \\
D & \leq & \min_{z(u)} \mathbf{E} \; d(X,z(U)).
\end{array}
\right\}.
\end{eqnarray*}

\begin{theorem}
\label{theorem main result}
\begin{eqnarray*}
{\cal S} & = & {\cal S}_0.
\end{eqnarray*}
\end{theorem}

Notice that the bound on $R$ does not depend on the choice of auxiliary variable $U$.  Quite intuitively, the minimum description rate needed is the entropy rate of the source $H(X)$.  This means that there is not a tradeoff between secret key usage and communication over the public channel.  The slice of the secrecy rate-distortion region that corresponds to a fixed value of $D$ is a rectangular region.

\section{Converse}
\label{section converse}

\begin{proof}
In this section we prove that ${\cal S} \subset {\cal S}_0$.  Therefore, we start by assuming that $(R_0,R,D)$ is in the interior of ${\cal S}$.  In other words, assume that rates $(R_0,R)$ can force distortion $D$.  Now we analyze the sequence of codes that achieves this. \footnote{The boundary of ${\cal S}$ is taken care of because ${\cal S}_0$ is a closed set.}

We start by bounding $R_0$ using Fano's inequality.
\begin{eqnarray*}
nR_0 & = & H(\omega) \\
& \geq & H(\omega|I) \\
& \geq & I(X^n;\omega|I) \\
& = & H(X^n|I) - H(X^n|I,\omega) \\
& \geq & H(X^n|I) - n \epsilon_n \\
& = & \sum_{q=1}^n H(X_q|I,X^{q-1}) - n \epsilon_n \\
& = & n H(X_Q|I,X^{Q-1},Q) - n \epsilon_n,
\end{eqnarray*}
where $\epsilon_n$ goes to zero as $n$ goes to infinity due to Fano's inequality, and $Q$ is uniformly distributed on $[n]$ and independent of $X^n$.

It is beneficial to assign names to the collections of random variables to correspond with the definition of ${\cal S}_0$.
\begin{eqnarray*}
X & \triangleq & X_Q, \\
U & \triangleq & \{I,X^{Q-1},Q\}.
\end{eqnarray*}
Notice that $X_Q \sim p_0(x)$ since $Q$ is independent of $X^n$.

Therefore, to summarize the bound on $R_0$,
\begin{eqnarray*}
R_0 & \geq & H(X|U) - \epsilon_n.
\end{eqnarray*}

The bound on $R$ is straightforward from lossless source coding and will be omitted.

Finally, we can massage the definition of achievability into a bound on the distortion $D$.
\begin{eqnarray*}
D & \leq & \frac{1}{n} \sum_{q=1}^n \min_{z^{(q)} (i,x^{q-1})} \mathbf{E} \; d(X_q,z^{(q)}(I,X^{q-1})) \\
& = & \mathbf{E} \; \left[ \min_{z^{(q)} (i,x^{q-1})} \mathbf{E} \; \left( d(X_Q,z^{(Q)}(I,X^{Q-1})) | Q \right) \right] \\
& = & \min_{z (i,x^{q-1},q)} \mathbf{E} \; d(X_Q,z(I,X^{Q-1},Q)) \\
& = & \min_{z(u)} \mathbf{E} \; d(X,z(U)).
\end{eqnarray*}

Since $\epsilon_n$ goes to zero and the space of distributions is compact, the converse is proven.
\end{proof}

\section{Sketch of Achievability}

\begin{proof}
In this section we outline the proof that ${\cal S} \supset {\cal S}_0$.  Because we use standard techniques of covering and binning, the proof only necessitates an outline.

Assume that $(R_0,R,D)$ is in the interior of ${\cal S}_0$, and consider the validating choice of $p(u|x)$.  First use a communication rate of $R_1 = I(X;U) + \epsilon$ to specify the index of a $U^n$ sequence from a rate-distortion-like codebook.  We can refer to this as a covering codebook.  Then use rates $R_2 = R_0 = H(X|U) + \epsilon$ bits to send a randomly assigned bin index based on the sequence $X^n$ and the common randomness $\omega$.  The message $I$ is the combination of these two communication steps, with a total rate of $R = R_1 + R_2$.  From standard techniques we know that $X^n$ can be decoded with high reliability from $I$ and $\omega$.

The second part of the message, which is the hash value of the source sequence $X^n$, has been hidden from the eavesdropper because the hash value is based on common randomness equal to the length of the message.  This is essentially an unstructured one-time-pad, and indeed a standard one-time-pad could very well be used instead for this part of the communication.

We now refer to the derivations in \cite{Cuff08} to claim that from the point of view of the eavesdropper, who knows only the sequence $U^n$ specified by the encoder, the posterior distribution of the source $X^n$ is close in total variation to the output of a memoryless channel from $U^n$ to $X^n$.  In other words,
\begin{eqnarray*}
p(x^n|i) & \approx & \prod_{j=1}^n p'(x_j|u_j(i)),
\end{eqnarray*}
where $p'(x|u) = \frac{p_0(x)p(u|x)}{\sum_{\tilde{x}} p_0(\tilde{x})p(u|\tilde{x})}$.

Given this posterior distribution of $X^n$, the eavesdropper minimizes the distortion by choosing the function $z(u)$ that minimized $\mathbf{E} \; d(X,z(U))$ and constructing the sequence $Z_i = z(U_i)$.  The resulting distortion is asymptotically
\begin{equation*}
\min_{z(u)} \mathbf{E} \; d(X,z(U)).
\end{equation*}
\end{proof}

\section{Linear Program Characterization}

The information theoretic region ${\cal S}_0$ accurately specifies the secrecy rate-distortion region; however, the optimization involved in this characterization yields an even simpler structure.  Due to the concave nature of the entropy function, the only efficient distributions $p(x|u)$ are corner points in the max-min distortion curve, which are points for which the eavesdropper can choose several different reconstructions $z_1, z_2, ...$ that each equally minimize the expected distortion.  This is characterized by the following set of distributions.
\begin{eqnarray*}
{\cal P} & \triangleq & \left\{ p(x) \; : \; \left| arg \max_z \mathbf{E} [d(X,z)] \right| \geq |{\cal X}| \right\},
\end{eqnarray*}
where $|{\cal X}|$ is the cardinality of the support of $X$ under the distribution $p(x)$.

As long as $d(x,z)$ is not redundant, meaning that there do not exist distinct $z_1$ and $z_2$ for which $d(x,z_1)=d(x,z_2)$ for all $x$, then ${\cal P}$ will be a finite set of points.  We assume this without loss of generality.  Let us refer to the distributions in ${\cal P}$ as vectors $p_1,...,p_K$, so that
\begin{eqnarray*}
{\cal P} & = & \{ p_1,...,p_K \}.
\end{eqnarray*}

Each point in ${\cal P}$ has an associated entropy
\begin{eqnarray*}
\alpha_k & \triangleq & H(p_k)
\end{eqnarray*}
and incurs a minimum distortion
\begin{eqnarray*}
\beta_k & \triangleq & \min_z \mathbf{E}_{p_k} \; d(X,z).
\end{eqnarray*}

\begin{theorem}
\label{theorem linear program}
The secrecy rate-distortion region can be characterized with a linear program.  Let $R = H(X)$ and $R_0$ be arbitrary.  Define $D^*$ to be the minimum $D$ such that $(R,R_0,D) \in {\cal S}$.  Then $D^*$ can be computed from the following linear program.
\begin{eqnarray*}
\mbox{maximize } & & \mu^T \beta \\
\mbox{subject to } & & \mu \in \Re_+^K, \\
& & \mu^T \alpha \leq R_0, \\
& & \sum_{k=1}^K \mu_k p_k = p(x).
\end{eqnarray*}
\end{theorem}

\section{Binary Source with Hamming Distortion}
\label{section binary}

Consider an information source that is an i.i.d. sequence of Bernoulli-$p$ bits (for $p<1/2$), with Hamming distortion incurred by the eavesdropper.  The applications of Theorems \ref{theorem main result} and \ref{theorem linear program} yield the following communication tactic.  First, use lossy compression to reveal the location of $(1-2p)$ zeros, leaving an equal number of ones and zeros concealed in the sequence.  When done carefully, the posterior distribution of the concealed bits will be consistent with unbiased i.i.d. coin flips.  Using the secret key, encrypt as many of these bits as possible.

With this technique, the distortion incurred will be $D = R_0/2$ for $R_0<2p$.  Notice that with perfect secrecy, created by applying a one-time-pad to the entire data at a secret key rate of $R_0 = H(p)$, the distortion incurred by the eavesdropper is $D=p$, since the eavesdropper will construct $z_i=0$ for all $i$.  We see that with optimal secret key usage, the full distortion $D=p$ can be forced on the eavesdropper while only using a secret key rate of $R_0 = 2p < H(p)$.  In other words, some reduction in secret key usage, below the requirement for perfect secrecy with a one-time pad, goes without penalty in this framework.  Furthermore, were we to use the secret key in a naive way and hide the first portion of the information sequence while revealing the rest, the resulting minimum distortion would be strictly suboptimal.

\bibliographystyle{IEEEtran}
\bibliography{ref}

\begin{thebibliography}{10}
\providecommand{\url}[1]{#1}
\csname url@samestyle\endcsname
\providecommand{\newblock}{\relax}
\providecommand{\bibinfo}[2]{#2}
\providecommand{\BIBentrySTDinterwordspacing}{\spaceskip=0pt\relax}
\providecommand{\BIBentryALTinterwordstretchfactor}{4}
\providecommand{\BIBentryALTinterwordspacing}{\spaceskip=\fontdimen2\font plus
\BIBentryALTinterwordstretchfactor\fontdimen3\font minus
  \fontdimen4\font\relax}
\providecommand{\BIBforeignlanguage}[2]{{%
\expandafter\ifx\csname l@#1\endcsname\relax
\typeout{** WARNING: IEEEtran.bst: No hyphenation pattern has been}%
\typeout{** loaded for the language `#1'. Using the pattern for}%
\typeout{** the default language instead.}%
\else
\language=\csname l@#1\endcsname
\fi
#2}}
\providecommand{\BIBdecl}{\relax}
\BIBdecl

\bibitem{ShannonSecrecy49}
C.~Shannon, ``Communication theory of secrecy systems,'' \emph{Bell Systems
  Technical Journal}, vol.~28, pp. 656--715, Oct. 1949.

\bibitem{Diffie76newdirections}
W.~Diffie and M.~Hellman, ``New directions in cryptography,'' \emph{IEEE Trans.
  on Info. Theory}, vol.~22, no.~6, pp. 644 -- 654, nov 1976.

\bibitem{BennettBrassard84}
C.~Bennett and G.~Brassard, ``Quantum cryptography: Public key distribution and
  coin tossing,'' in \emph{IEEE International Conference on Computers, Systems,
  and Signal Processing}, Bangalore, India, Dec. 1984.

\bibitem{WynerWiretap75}
A.~Wyner, ``The wire-tap channel,'' \emph{Bell Systems Technical Journal},
  vol.~54, pp. 1355--1387, Oct. 1975.

\bibitem{4594958}
D.~Gunduz, E.~Erkip, and H.~Poor, ``Lossless compression with security
  constraints,'' in \emph{IEEE International Symposium on Information Theory},
  6-11 2008, pp. 111 --115.

\bibitem{Shannon48}
C.~Shannon, ``A mathematical theory of communication,'' \emph{Bell Systems
  Technical Journal}, vol.~27, pp. 379--423,623--656, July and Oct. 1948.

\bibitem{1176626}
N.~Merhav, ``A large-deviations notion of perfect secrecy,'' \emph{IEEE Trans.
  on Info. Theory}, vol.~49, no.~2, pp. 506 --508, feb. 2003.

\bibitem{Maurer93}
U.~Maurer, ``The role of information theory in cryptography,'' in \emph{4th IMA
  Conference on Cryptography and Coding}, Cirencester, England, Dec. 1993.

\bibitem{YamamotoCipherDistortion97}
H.~Yamamoto, ``Rate-distortion theory for shannon cipher systems,'' \emph{IEEE
  Trans. on Info. Theory}, vol.~43, no.~3, pp. 827--835, May 1997.

\bibitem{Cuff08}
P.~Cuff, ``Communication requirements for generating correlated random
  variables,'' in \emph{IEEE International Symp. on Info. Theory}, Toronto,
  2008, pp. 1393--1397.

\end{thebibliography}

\end{document}